\magnification1200


\vskip 2cm
\centerline
{\bfÊ On the different formulations of the  E11 equations of motion}
\vskip 1cm
\centerline{ Peter West}
\centerline{Department of Mathematics}
\centerline{King's College, London WC2R 2LS, UK}
\vskip 2cm
\leftline{\sl Abstract}
The non-linear realisation of the semi-direct product of $E_{11}$ with its vector representation leads toÊ  equation of motions for the fields graviton, three form, six form, dual graviton and the level four fields which correctlyÊ describe the degree of freedom of eleven dimensional supergravity at the linearised level. The equations with one derivative generically hold as equivalence relations  and are often duality relations. From these equations, by taking derivatives,  one can derive equations that are equations of motion of the  familiar kind. The entire hierarchy of equations is $E_{11}$ invariant and the  constructionÊ does not require any steps  beyond $E_{11}$.  We review these past developments with an emphasis on the featuresÊ that have been overlooked in hep-th:1703.01305,  whose alternative approach we also comment on.  
\vskip2cm
\noindent

\vskip .5cm

\vfill
\eject

\medskipÊ
{\bf 1. Introduction}
\medskip

We consider theÊ non-linear realisation of the semi-direct product ofÊ $E_{11}$ and its vector representation, denoted $E_{11}\otimes_s l_1$ [1,2]. For a review of this subject see reference [3].Ê In this paper weÊ will focus in the eleven dimensional theory which arises when one takes the decomposition of $E_{11}$ into the GL(11) subalgebra that arises from deleting the nodeÊ usually labelled as eleven inÊ the $E_{11}$ Dynkin diagram. One finds that this theory contains the fields [1,4]
$$
h_a{}^b ;\ A_{a_1a_2a_3} ;\ A_{a_1\ldots a_6} ;\ h_{a_1\ldots a_8,b}; \ 
A_{a_1\ldots a_9, b_1b_2b_3}, \ Ê    A_{a_1\ldots a_{10}, b_1 b_2} , \ Ê
A_{a_1\ldots a_{11}, b} , \ldotsÊ
\eqno(1.1)$$
The first four fields correspond to theÊ graviton, the three form, the six form and the dual graviton field at levels zero, one,Ê two and three respectively. For these fieldsÊ theÊ indices in a given block are antisymmetric and the dual gravitonÊ field obeys the constraint [1,4]
$$
h_{[a_1\ldots a_8,b ]}=0Ê
\eqno(1.2)$$
The last three fields explicitly listed are thoseÊ at level four and we will give their constraints later in the paper.Ê
The non-linear realisation contains a spacetime with the coordinates [2]
$$
x^a,\  x_{ab},  \  x_{a_1\ldots a_5}, \  x_{a_1\ldots a_7,b},Ê\  x_{a_1\ldots a_8},\ldotsÊ Ê
\eqno(1.3)$$
and the fieldsÊ depend on these coordinates.
\par Ê
The equations of motion are determined by the symmetries of the non-linear realisationÊ whose procedure can be found in many $E_{11}$ papers, see for example [3,6]. While partial results can be found in the early $E_{11}$ papers it is only relatively recently thatÊ a systematic study of the equations of motion  have been carried out  using the higher level $E_{11}$ symmetries andÊ a better understanding of some of the more intricate details of how the non-linear realisation works in practice.Ê
It wasÊ shown [5,6] that the non-linear equations of motion of the eleven dimensional theory are uniquely determinedÊ at low levels and they are precisely those of eleven dimensional supergravity of Cremmer, Julia and Scherk when suitably truncated.Ê The analogous calculations have also been carried out in five dimensions [5,6]Ê with the same result. It is inevitable that similar results apply in all the other dimensions. This essentially confirms the $E_{11}$ conjecture, namely that the low energy effective action of strings and branes has an $E_{11}$ symmetry. These results were obtained at the full non-linear level and they included all the non-linear effects in the corresponding supergravity theories.Ê
\par
The results were extended to a higher level to include the level four fields, albeitÊ at the linearised level [7]. The $E_{11}$ invariant equations of motion of  these fields were found. One findsÊ that the equations correctly account for all the degrees of freedom of eleven dimensional supergravity through a series of duality relations as well as giving the eleven dimensional origin of Romans theory [7,8].Ê
\par
In a very recent paper [9] the authors start from the $E_{11}$ formalism and also compute the equations of motion, at the linearised level,Ê along the lines of some of the earlier $E_{11}$ papers [1,2]. However, they have overlooked an important aspect of how the duality relations work and have concluded that the trace of the spin connection in missing and as a result  the theory does not describe Einstein's theory. In this paper we will review the work on $E_{11}$ of references [5,6,7] emphasising the points that has been overlooked  in [9], namely that the equations that have low numbers of derivatives generically hold as equivalence relations. While  the equations of motion with the highest number of derivatives, for a given field,  hold in the usual sense and they are the correct equations of motion in a familiar form [10,5,6,7].Ê

\medskip
Ê{\bf 2. Equations of motion including level three}
\medskip
The construction of the equations of motion from the $E_{11}\otimes_s l_1$ non-linear realisation uses the Cartan forms which transform only under the action of the Cartan involution invariant subalgebra of $E_{11}$, denoted $I_c(E_{11})$,  once one converts its one world index to a tangent index.Ê TheÊ set of equations is essentially unique, at least at low levels, and isÊ invariant under the symmetries of the non-linear realisation. One findsÊ a hierarchy of equations and the results up to and including level three fields are given in the table below [5,6,7].Ê The extension to level four will be reviewedÊ later in thisÊ paper. As one movesÊ down the table the number of derivatives increases, indeed the superscript in brackets denotes the number of derivatives, and as one moves to the right the level of the fields involved increases.ÊThe horizontal arrows denote the effect of $I_c(E_{11})$ transformations. 

\bigskip
{\bf Table 1. The $E_{11}$ variations of the equations of motion up to level three }
\medskip
$${\normalbaselineskip=20pt \matrix{Ê
E^{(1)}{}_{a_1a_2a_3a_4}=0 & \Leftrightarrow & E^{(1)}{}_a{}^{b_1b_2} \dot =0 \cr
\Downarrow & ÊÊ& \Downarrow Ê& ÊÊ \crÊ
E^{(2)}{}_{a_1a_2a_3}=0 Ê& ÊÊ& E^{(2)}{}_a{}^b=0 Ê& Ê\cr
& Ê\Leftrightarrow & Ê&Ê \cr
ÊE^{(2)}{}_{a_1\ldots a_6}=0 Ê& ÊÊ& ÊE^{(2)}{}_{a_1\ldots a_8,b} =0 & \cr
\cr}}
$$
\bigskip
The objects that appear inÊ the above tableÊ with one derivative are given byÊ
$$
{E^{(1)}}_{a_1\ldots a_4}\equivÊ \partial_{[a_1} A_{a_2a_3a_4]}
-{1\over 48}\epsilon _{a_1a_2a_3a_4}{}^{b_1\ldots b_7} \partial_{b_1}A_{b_2\ldots b_7 }Ê =0
\eqno(2.1)$$
$$
{ E}^{(1)}{}_{a,\,b_1 b_2} \equiv \omega_{a,\,b_1b_2} - {1\over 4}\,\varepsilon_{b_1b_2}{}^{c_1...c_9}\,\partial_{c_1} h_{c_2...c_9,\,a} \dot =0
\eqno(2.2)$$
whereÊ
$$
\omega _{a, bc} = - \partial_{b}h_{ (ca)}+ \partial_{c}h_{ (ba)}+ \partial_{a}h_{ [bc]}
\eqno(2.3)$$
While the equationsÊ with two derivatives are given byÊ
$$
E^{(2)}{}^{a_1a_2a_3}\equiv \partial_{b}E^{(1)}{}^{ba_1a_2a_3} = \partial_{b}
\partial^{[b} A^{a_1a_2a_3]} =0
\eqno(2.4)$$
$$
{E^{(2)}}{}^{a_1\ldots a_6}\equiv {2\over 7!} \partial_{b}\epsilon ^{bÊ a_1\ldots a_6 c_1\ldots c_4 }E^{(1)}{}_{c_1\ldots c_4}=
\partial_{b}\partial^{[b} A^{  a_1\ldots a_6]}=0
\eqno(2.5)$$
$$
E^{(2)}{}_a{}^b= \partial_a \omega_{c,}{}^{bc}-\partial _c \omega _{a, }{}^{bc} \equiv R_a{}^b=0
\eqno(2.6)$$
and Ê
$$
E^{(2)}{}_{a_1\ldots a_8,}{}^{ b}\equivÊ
-\,{1\over 4}\,\partial^{[d}\,\partial_{[d}h_{\,a_1...a_8],\,}{}^{b]}=0
\eqno(2.7)$$

In the above objects we have only included the parts that contain terms withÊ derivatives with respect to the usual spacetime coordinates, the terms with derivatives with respect to the level one coordinates have been found in [7]. However, it is important to realise that the latter terms are absolutely crucial for the invariance of the equations and we have omitted them here in order to make the presentation uncluttered and  clearer. In the previous papers [5,6,7], the equations of motion were formulated in terms of the Cartan forms, to recover these results one makes the substitution $\partial_a A_\star \to G_{a, \star}$ which is valid at the linearised level.Ê
\par
In the above table the equations with only one derivative areÊ duality relations, that is they express the derivative of one field in terms of the derivative of another field and they appear in the first row.Ê One can take a derivative of these equations with one spacetime derivative and find equations that are second order in derivatives. If one does this in a precise way one can Ê eliminate one of the two fields that occur in the equation with one spacetime derivative and find an equation of motion for only one of the fields that is second order in derivatives.Ê These are the equations in the second row. The down arrow ($\Downarrow$) in the table correspond to this projection using the derivative. These later equationsÊ are theÊ equations of motion forÊ the degrees of freedom of eleven dimensional supergravity. However, they appear in a duality symmetric formulation in that one find equations in their usual form, that is for the graviton (2.6) and three form (2.4) as well as equivalent equations for the dual fields, that is for the six form (2.5) and the dual graviton (2.7).Ê It isÊ important to realise that one can not discard the first order equations as they ensure that the degrees of freedom are not duplicated.Ê
\parÊ
We note that in general the precise form of the equations of motion for a given particle are determined by the requirement that they describe the corresponding irreducible representations of the Poincare group. The one exception being thatÊ one can use fields of different Lorentz character to describe the same particle.Ê We note that the correct degrees of freedom of gravity are described by a dual graviton field $h_{a_1\ldots a_8 ,b}$ that has the equation of motion of equation (2.7) and alsoÊ that obeys the constraint of equation (1.2). As one would expect, the equation of motion for the dual gravitonÊ obeys the same constraint, that is, the equationÊ obeys the condition $E^{(2)}{}_{[ a_1\ldots a_8,}{}_{ b]}=0$ [7]. Thus gravity can be  formulated in terms of a dual field that satisfies equation (1.2) and the field equation has the corresponding constraint. Put differently it does not also need a field that has nine indices which are totally antisymmetric.Ê
\par
Equation (2.1) is the well know duality relation between the three form and the six form. WhileÊ equation (2.2)Ê relates the graviton field to the dual graviton field. This equation was given in the original $E_{11}$ paper [1] {\bf with the difference that the constraint of equation (1.2) was not enforced} as the equation was not derived from $E_{11}$ but rather constructed by hand in order to show that the field $h_{a_1\ldots a_8,b}$ could correctly describe the degree of freedom of gravity.Ê
In fact that such a field did describe gravity was first realised in five dimensions in reference [11] and such a field was suggested in a general dimension in reference [12],  although only partialÊ light cone based arguments were given to show that it really did describe gravity. It was shown in reference [22] that the equations of reference [1] were equivlent to the formulation of reference [11] in five dimensions. Thus theÊ equation in reference [1] includedÊ a field $h_{a_1\ldots a_9}$, that is,Ê the completely antisymmetric part of $h_{a_1\ldots a_8,b}$. As equation (1.2) makes clear this totally antisymmetric field isÊ not contained in the $E_{11}$ non-linear realisation. It wasÊ realised in reference [4] that a version of theÊ gravity-dual gravity equation (2.2),Ê which includes the field $h_{a_1\ldots a_9}$ and had terms with the trace of the spin connection (see equation (4.19) of reference [1] or equation (4.3) of reference [4])was Lorentz invariant if the usual Lorentz transformation of the graviton was compensated for by a transformation of the nine form indeedÊ
$$
\delta h_{ab} =-\Lambda _{ab}Ê ,\quad \deltaÊ h_{a_1\ldots a_{9}}= {1\over 4.7!}Ê
\epsilon _{a_1\ldots a_{9}  c_1c_2}\Lambda ^{c_1c_2}Ê
\eqno(2.8)$$
This transformations follows in an obvious way once one takes into account the well known transformation ofÊ Ê theÊ spin connection whichÊ transforms by the inhomogeneous term,Ê $\delta \omega_{\lambda,\,\mu_1\mu_2} = \partial_\lambda \Lambda _{\mu_1\mu_2}+\ldots $ where $+\ldots $ indicate the homogeneous terms.  
\par
Now let us consider the {\bf gravity-dual gravity equation (2.2) which appears in the $E_{11}$ non-linear realisation where the dual graviton satisfies equation (1.2)}.Ê
Clearly this equation is not invariant under local Lorentz transformations. The resolution of this dilemma is to think of equation (2.2) as beingÊ Ê valid only modulo local Lorentz transformations. In other words it holds modulo the transformationsÊ [10,5,6,7]
$$
E^{(1)}{}_{a,b_1b_2}\sim E^{(1)}{}_{a,\,b_1 b_2} +\partial_{a} \Lambda_{b_1b_2}+\ldotsÊ
\eqno(2.9)$$
where $+\ldots $ indicate the homogeneous Lorentz transformations of $E^{(1)}{}_{\lambda,\,\mu_1\mu_2}$.Ê The use of the symbol $\dot =$ in equation (2.2) implies that the equation only holds modulo the local Lorentz transformations as just discussed.Ê
\par
\par
To put it in aÊ more mathematical sense;Ê we are regarding equation (2.2) to belong to Ê an equivalence class, the equivalence relation being that of equation (2.9). Physicists are familiar to such relations, for exampleÊ in the definition of physical states in the context of the BRST formalism.Ê This way of proceeding is a completely correct and rigorous. As one takes the derivatives to find the second order equations one eliminates theÊ transformations that the equation is modulo, indeed the projection is chosen in just such a way as to do this, and one finds equations which hold in the usual sense, that is, are not modulo any transformations.Ê
\par
We note that taking equation (2.2) to hold as an equivalence relation has the same effect as adding a field $h_{a_1\ldots a_9}$ and then realising that the equation is invariant under the Lorentz transformations of equation (2.8). In effect the Lorentz transformation is just the field $h_{a_1\ldots a_9}$.Ê
\par
It was realised in references [13] that if one naivelyÊ takes the trace of equation (2.2)Ê then the trace of the spin connection vanished due to the condition of equation (1.2) andÊ we find thatÊ
$$
E^{(1)}{}_{c,}{}^{c b}=\omega_{c ,}{}^{c b}=Ê \partial^{b} h^c{}_c - \partial_c h^{c b} =0
\eqno(2.10)$$
Clearly this is not consistent with Einstein's equation.Ê
However, as we haveÊ discussed just above,Ê equation (2.2) is to be thought of as an equivalence relation and we have toÊ take this into account. Following reference [10],Ê and its implementation in references [5,6,7],Ê one find that if we take the traceÊ of equation (2.2) we find,Ê instead of equation (2.10),Ê the result
$$
E^{(1)}{}_{c ,}{}^{cb}=\partial^{b} h^c{}_c - \partial_c h^{(bc)} + \partial_c k^{cb}=0
\eqno(2.11)$$
where $k_{ab}=(h_{[ab]}- \Lambda _{ab} )$ and $\Lambda _{ab}$ is the Lorentz transformation the equation is subject to. Taking the derivative of equation (2.2) we find thatÊ
$$
Ê\partial^{c} E^{(1)}{}_{a,cb}=Ê
\partial_{b}\partial^{c} h_{(c a)}- \partial^2h_{(ba )} Ê
+\partial_{a} \partial^{c} k_{cb}=0Ê
\eqno(2.12)$$
where $\partial^2= \partial^c\partial_c$.

To eliminate the local Lorentz transformation we can act on equation (2.11) with $\partial_{a}$ and subtractÊ equation (2.12),Ê to find an equation that is Ê independent of the dual graviton and the local Lorentz transformation andÊ is given byÊ
$$
Ê\partial^{c} E^{(1)}{}_{a,cb}- \partial_{a} E^{(1)}{}^c{}_{,c b}= \partial _{a} \partial^c h_{(cb)}+ \partial_{b} \partial^c h_{(ca)} - \partial^2 h_{(a b)} - \partial_{b }\partial_{a} h_c{}^c=0
\eqno(2.13)$$
This is indeed the linearised Einstein equation.
\par
We observe that acting with the  derivative $\partial _{b}$Ê on equation (2.11)Ê weÊ can also find an equation that is independent of the local Lorentz transformationÊ
$$
\partial^2 h^b{}_b - \partial ^{a} \partial^b h_{a b}=0
\eqno(2.14)$$
which is indeed the linearised version of the trace of the Einstein equation.Ê
\parÊ
The authors of reference [9] have studied reference [13] and so are aware of the difficulties on taking the trace of equation (2.2) and the contradiction embodied in equation (2.10). However, they have not realisedÊ theÊ corresponding resolution of the difficulties [10, 5,6,7 ] and so come to the conclusion that one must add fields. ÊUnfortunately reference [13] did previously  realise that the gravity-dual gravity equation was an equivalence relation but then this paper incorrectly concluded that this did not solve the problem whereas, as we have just seen,  it does [10,5,6,7]. 
\par
To summarise, the non-linear realisation of $E_{11}\otimes_s l_1$ leads to equations that do not miss the trace of the spin connection as the equation in which it occurs is an equivalence relation. From this equation one finds the correct equation for the degrees of freedom of the gravity whetherÊ encoded in the usual graviton field, or the dual gravity field. Ê
\par
The above story is the general pattern; the equations of motion with low numbers of derivatives are equivalence relations and from these one can derive, by taking derivatives, equations that hold in the traditional sense and these are the familiar equations of motion for the fields involved. The full system of equations is invariant under the symmetries of the non-linear realisation once one takes account of the fact that some of the equations are equivalence relations.Ê
In the $E_{11}\otimes_s l_1$ non-linear realisation we find fields with more and more blocks of indices asÊ the level increases and one can expect that these should obey the traditional equations of motion that have one derivative for every block of indices that they contain. Hence as one considers higher and higher level fields in $E_{11}$ oneÊ expects to find traditional equations with higher andÊ higher numbers of derivatives. The way $E_{11}$ handles this situation is to have a hierarchy of equationsÊ containing increasing numbers of derivatives. The equations with the lower number of derivatives are equivalence relations, that is, only hold modulo certain transformations.  The very first duality relation of equation (2.1) is unusual in that it already possess one derivative for each block of indices that the field carries, namely one. This pattern becomes particularly apparent when one considers level four fields as we do in the next section.Ê
\par
Reference [14] established a no-go theorem for the dual graviton and this is often quoted as an obstacle to the $E_{11}$ programme.Ê
As we have seen above there is no obstacle at the linear level and one finds a perfectly correct theory. These is also no obstacle at the non-linear level as reference [14] investigates if one can find a theory that involves the dual graviton field alone. However, this is not the path chosen by $E_{11}$ which involves in the non-linear dual graviton equation both the gravity and the dual gravity fields. This equation is under construction and will be published elsewhere [15].Ê
\par
The transformations of the equations of motion given in table 1 were given in reference [7] and for completeness we give the results here. The horizontal arrows correspond to the effect of varying under $I_c(E_{11})$ which is the non-trivial symmetry which acts on the Cartan forms which contain the above objects. The transformations have a parameter $ \Lambda_{ b_1 b_2 b_3}$, for an explanation see the earlier papers [3,4,5,6,7,13] . WeÊ begin with the transformations of the duality relations with only one derivative:
$$
\delta {\cal Ê E}^{(1)}{}_{a_1\ldots a_4}= {1\over 4!} \epsilon _{a_1\ldots a_4 }{}^{b_1\ldots b_7} \Lambda_{b_1 b_2b_3}E^{(1)}{}_{b_4 \ldots b_7}+Ê
3E^{(1)}{}_{c ,}{}_{[a_1a_2} \Lambda ^c{}_{a_3a_4]}
\eqno(2.15)$$
$$
\delta{\cal E}^{(1)}{}_{\lambda,\,\mu_1\mu_2} = {7\over 12}\,\varepsilon_{\mu_1\mu_2}{}^{\nu_1...\nu_6\sigma_1\sigma_2\sigma_3}\, E^{(1)}{}_{\lambda\nu_1...\nu_6}\,\Lambda_{\sigma_1\sigma_2\sigma_3} + {1\over 2}\,\varepsilon_{\mu_1\mu_2}{}^{\nu_1...\nu_7\sigma_1\sigma_2}\, E^{(1)}{}_{\nu_1...\nu_7}\,\Lambda_{\sigma_1\sigma_2\lambda}
$$
$$
+{55\over 2} \Lambda_{\sigma_1\sigma_2 [\mu_1}\epsilon _{\mu_2 ] }{}^{\nu_1\ldots \nu_{10} } E^{(1)}{}_{\nu_1\ldots \nu_{10} , \lambda}{}^{\sigma_1\sigma_2}
-{55\over 18}Ê \Lambda^{\sigma_1\sigma_2 \sigma_3}\eta_{\lambda[\mu_1} \epsilon _{\mu_2]}{}^{ \nu_1\ldots \nu_{10} } E^{(1)}{}_{\nu_1\ldots \nu_{10} , \sigma_1\sigma_2\sigma_3} Ê
$$
$$
+ {3\over 4}\,\Lambda_{\mu_1\mu_2}{}^{\sigma}\,\varepsilon^{\rho_1...\rho_{11}}\,E^{(1)}{}_{\rho_1,\,\rho_2\ldots \rho_{11},\sigma\lambda}
+\partial_\lambda \tilde \Lambda_{\mu_1\mu_2}Ê
\eqno(2.16)$$
whereÊ
$$
\partial_\lambda \tilde \Lambda_{\mu_1\mu_2}=Ê
-\,\varepsilon_{\mu_1\mu_2}{}^{\nu_1...\nu_9}\,\,\Bigg[{1\over 12}\,\partial_{\lambda}A_{\nu_1...\nu_6}\,\Lambda_{\nu_7\nu_8\nu_9} 
$$
$$
+ {55\over 36}\,\partial_{\lambda}A_{\nu_1...\nu_9,\,\sigma_1\sigma_2\sigma_3}\Lambda^{\sigma_1\sigma_2\sigma_3}
 + {55\over 16}\,\partial_{\lambda}A_{\sigma_1\sigma_2\sigma_3\nu_1...,\,\nu_9}\Lambda^{\sigma_1\sigma_2\sigma_3}\Bigg],
\eqno(2.17)$$
The transformations include the contributions from level four fields which we will discuss in the next section. We observe that the variation of the gravity-dual gravity relations involves a local Lorentz transformation consistent with the fact that this relation is an equivalence relation. The symbol ${\cal E}$  is equal to $E$ plus terms that  contain the derivatives with respect to the higher level coordinates. The precise form of the ${\cal E}$'s can be found in reference [7] and we note that without these the result would not hold.Ê
\par
The variations of the equations of motion with two spacetime derivatives are given by
$$
\delta{\cal E}^{(2)}{}^{a_1a_2a_3 }= {3\over 2} E^{(2)}{}_b{}^{[a_1|}\Lambda ^{b | a_2a_3]}Ê
-{1\over 24} \epsilon ^{a_1 a_2 a_3 \nu c_1\ldots c_4 b_1b_2b_3}\partial_\nuÊ E^{(1)}{}_{c_1\ldots c_4} \Lambda_{ b_1 b_2 b_3}
$$
$$
= {3\over 2} E^{(2)}{}_b{}^{[a_1|}\Lambda ^{b | a_2a_3]} +3.5.7E^{(2)}{}^{a_1a_2a_3 b_1b_2b_3} \Lambda_{b_1b_2b_3}
Ê\eqno(2.18)$$
$$
\delta {\cal E}^{(2)}{}_{a_1\ldots a_6}= {8\over 7}\Lambda_{[a_1a_2a_3}E^{(2)}{}_{a_4a_5a_6]}Ê
- 27.64 E^{(2)}{}_{a_1\ldots a_6 c_1c_2, c_3} \Lambda^{c_1c_2c_3}
\eqno(2.19)$$

$$
\delta {\cal E}^{(2)}{}_{ab} = -36 \Lambda ^{d_1d_2}{}_{ a} E^{(2)}{}_{bd_1d_2} -36 \Lambda ^{d_1d_2}{}_{ b} E^{(2)}{}_{ad_1d_2} +8\eta_{ab} E^{(2)}{}_{d_1d_2d_3}\Lambda ^{d_1d_2d_3}
\eqno(2.20)$$

$$
\delta{\cal E}^{(2)}{}_{\,\rho_1...\rho_8,\,\lambda} = -\,{7\over 4}\,E^{(2)}{}_{\sigma [\rho_1...\rho_5 }\,\Lambda^{\sigma }{}_{\rho_6\rho_7}\eta_{\rho_8]\lambda}\,
$$
$$
+\,275\,\left({ E^{(2)}{}}_{\rho_1...\rho_8\sigma_1,\,\sigma_2\sigma_3\lambda} - {1\over 9}\,{ E^{(2)}{}}_{\rho_1...\rho_8\lambda,\,\sigma_1\sigma_2\sigma_3}\right)\,\Lambda^{\sigma_1\sigma_2\sigma_3}
$$
$$
+{1\over 4.7!}\left(\epsilon_{\rho_1\ldots \rho_8 \sigma_1\tau_1\tau_2}\partial^{\tau_1} E^{(1)}{}^{\tau_2 \sigma_2\sigma_3 \lambda}
-{1\over 9} \epsilon_{\rho_1\ldots \rho_8 \lambda\tau_1\tau_2}\partial^{\tau_1} E^{(1)}{}^{\tau_2 \sigma_1\sigma_2\sigma_3} 
\right)
$$
$$
+\,{165\over 8}\,\left(E^{(2)}{}_{\nu \rho_1...\rho_8\sigma_1\sigma_2, \nu\lambda , \sigma_3}Ê - {1\over 9}\,E^{(2)}{}_{\sigma_2\rho_1...\rho_8\lambda\nu, \sigma_1\sigma_3 , \nu}\right)\,\Lambda^{\sigma_1\sigma_2\sigma_3}
\eqno(2.21)$$
In the above we  have included the level four terms whose definitions are given in the next section. We observe  that the equations do rotate into each other under the symmetries of the non-linear realisation provided we take account of the fact that one of them is an equivalence relation.


\medskip
{\bf 3. Equations of motion at level four}
\medskip
The fields of the non-linear realisation given in equation (1.1) include those of level four and for these fields allÊ blocks of indices are antisymmetrised except for Ê the second block ofÊ the field $A_ {a_1\ldots a_{10}, b_1b_2}$ which is symmetric, that is, $A_ {a_1\ldots a_{10}, b_1b_2}= A_ {a_1\ldots a_{10}, b_2b_1}$. The fields also obey the usual SL(11) irreducibility conditions, that is,Ê
$$
A_{[ a_1\ldots a_9, b_1 ] b_2 b_3}=0 = Ê A_ {[ a_1\ldots a_{10}, b_1 ]b_2} 
\eqno(3.1)$$.Ê
\par
The equations of motion,  including the above fields,  are given in the table 2 below,  so  extending  the results of the previous section which were  given in table 1.Ê
\bigskip
{\bf Table 2. The $E_{11}$ equations of motion including level four fields} [7].Ê
\medskip
$${\normalbaselineskip=20pt \matrix{Ê
E^{(1)}{}_{a_1a_2a_3a_4}=0 & \Leftrightarrow & E^{(1)}{}_a{}^{b_1b_2} \dot =0 & \Leftrightarrow & E^{(1)}{}_{a_1\ldots a_{10}, b_1b_2b_3}\dot =0 , \ E^{(1)}{}_{a_1\ldots a_{11}, b_1b_2}\dot =0\cr
\Downarrow & ÊÊ& \Downarrow Ê& ÊÊ& \Downarrow \crÊ
E^{(2)}{}_{a_1a_2a_3}=0 Ê& ÊÊ& E^{(2)}{}_a{}^b=0 Ê& Ê& E^{(2)}{}_{a_1\ldots a_9 , b_1b_2b_3}=0 Ê\cr
& Ê\Leftrightarrow & Ê& \Leftrightarrow & \cr
ÊE^{(2)}{}_{a_1\ldots a_6}=0 Ê& ÊÊ& ÊE^{(2)}{}_{a_1\ldots a_8,b} =0 & & ÊE^{(2)}{}_{a_1\ldots a_{11}, b_1b_2, c}\dot =0 \cr
& & & &\Downarrow \cr
& & & & E^{(3)}{}_{a_1\ldots a_{11}, b_1b_2, c_1 c_2}=0Ê \cr\cr}}
$$
\medskip
In the above table 2 theÊ objects not contained in table 1 with one derivative are given byÊ
$$
ÊE^{(1)}{}_{\mu_1\ldots \mu_{10} , \sigma_1\sigma_2\sigma_3}
\equiv \partial_{[\mu_1}A_{ , \ldots \mu_{10}] , \sigma_1\sigma_2\sigma_3}
-{1\over 5.5.11.7!} \epsilon _{\mu_1\ldots \mu_{10}}{}^{\tau} \partial_{[\tau} A_{ \sigma_1\sigma_2\sigma_3] }\dot =0
\eqno(3.2)$$
$$
ÊE^{(1)}{}_{\rho_1\,\rho_2\ldots \rho_{11},\sigma\lambda}\equiv \partial_{[\rho_1} A_{\rho_2\ldots \rho_{11}],\sigma\lambda}\dot =0 ,Ê
\eqno(3.3)$$
and those with two derivatives byÊ
$$
E^{(2)}{}_{\rho_1...\rho_9,\,}{}^{\sigma_1\sigma_2\sigma_3} \equiv E^{(2)}{}_{\nu\rho_1...\rho_9,\,}{}^{\nu\sigma_1\sigma_2\sigma_3} \dot =0
\eqno(3.4)$$
whereÊ
$$
E^{(2)}{}_{\mu_1...\mu_{10},\,\sigma_1...\sigma_4} \equiv \partial_{[\sigma_1 |}\,
E^{(1)}{}_{\mu_1\ldots \mu_{10} ,| \sigma_2\sigma_3\sigma_4 ]}
\eqno(3.5)$$
andÊ
$$
E^{(2)}{}_{\nu_1Ê \nu_2\ldots \nu_{11} , Ê \kappa\tau, \rho}Ê \equiv \partial _\tau E^{(1)}_{\nu_1 \nu_2\ldots \nu_{11},Ê \rho \kappa} - \partial _\kappa E^{(1)}_{\nu_1Ê \nu_2\ldots \nu_{11} ,Ê \rho \tau} \dot =0.
\eqno(3.6)$$
The one object with three derivatives is defined to beÊ
$$
ÊE^{(3)}{}_{c_1...c_{11},\,a_1a_2,\,b_1b_2} = -{1\over 2}(\partial_{a_1} E^{(2)}{}_{c_1...c_{11},\,b_1b_2,\,a_2} - \partial_{a_2}\, E^{(2)}{}_{c_1...c_{11},\,b_1b_2,a_1})
$$\
$$
= 2\,\partial_{[a_1}\,\partial_{[b_1}\,\partial_{[c_1}A_{c_2...c_{11}], a_2]b_2]}=0,
\eqno(3.7)$$
As before the down arrow means we take a derivative to find a new equation, the  precise projection is given in the above definitions.Ê The horizontal arrows in table 2 correspond to the $I_c(E_{11})$ variations which are discussed below. 
\par
As before the presence of the symbol $\dot =$ indicates that the equation should be viewed as an equivalence relation, that is equations (3.2), (3.3) and equation (3.6). We now comment on the general procedure for finding what are the transformations that are required in the equivalence relations. These can be found in two ways. Either by integrating up the exact equation in the hierarchy which has the largest number of derivatives, or by carrying out an $I_c(E_{11})$ transformation on the equation and finding out what additional transformations in addition to the  previously found equations arise. Of course having found the results using one method one can check the results using the other method. We now illustrate the two methods in the context of the $A_{a_1\ldots a_9 , b_1b_2b_3}$ field and the $A_{a_1\ldots a_{10} , b_1b_2}$ field respectively. 
\par
The equations that involve the three form and the 
$A_{a_1\ldots a_9 , b_1b_2b_3}$ field are those of equations  (3.2), (3.4)Ê and (3.5). The equation $E^{(2)}{}_{\rho_1...\rho_9,\,}{}^{\sigma_1\sigma_2\sigma_3} =0$ arises from the $I_c(E_{11})$ variation of the dual graviton equation of motion  (2.21). It isÊ
a duality relations between the three form $A_{a_1a_2a_3}$ and the field $A_{a_1\ldots a_9 , b_1b_2b_3}$. However, in order to eliminate the field $A_{a_1a_2a_3}$ in Ê equation (3.4) we must take the triple trace Ê to find the equationÊ
$$
E^{(2)}{}_{\rho_1...\rho_6 \nu_1\ldots \nu_4,}{}^{\nu_1\dots \nu_4} =Ê
\partial^{[\nu_1} \partial_{[\rho_1}A_{\rho_2 ..\rho_6 \nu_1\ldots \nu_4 ],}{}^{\nu_2\dots \nu_4]}=0
\eqno(3.8)$$
This is indeed the correct equation of motion for the $A_{a_1\ldots a_9, b_1b_2b_3}$ to describe the same degrees of freedom which are usually encoded in the three form [11]. In fact from the first order duality relation, equation (3.2) one can deduce a stronger condition, namely $E^{(2)}{}_{\mu_1...\mu_{10},\,\sigma_1...\sigma_4} =0$, rather than equation (3.4) even taking account of the modulo transformations. This later equation may be contained in the non-linear realisation as it could  also result from varying the equations of motion of the level five fields. 
\par
Equation (3.4), which has two derivatives,  can be integrated up to find a first order equation, namely equation (3.2), that involves the same fields. 
The result is equation (3.2) which involves the object $E^{(1)}{}_{a_1\ldots a_{10}, b_1b_2b_3}$ which is  first order in derivatives. ThisÊ equation is an equivalence relation which can be written asÊ
$$Ê
E^{(1)}{}_{\mu_1\ldots \mu_{10} , \sigma_1\sigma_2\sigma_3}\sim E^{(1)}{}_{\mu_1\ldots \mu_{10} , \sigma_1\sigma_2\sigma_3}+ \partial_{[\sigma_1 |}\partial_{[\mu_1}\Lambda_{\mu_2\ldots \mu_{10}] , |\sigma_2\sigma_3]}
\eqno(3.9)$$
How to integrate up the equation wasÊ discussed  in reference [16], on page 26. One would expect in this process to find a single derivative acting on a new function, however,  this new function can in turn be written as the derivative of anther  function $\Lambda_{\mu_2\ldots \mu_{10} , \sigma_2\sigma_3}$ as a result of  the fact that the field $A_{a_1\ldots a_9, b_1b_2b_3}$ is an irreducible representation of SL(11). We refer the reader to reference [16] for the complete discussion. One can also recover equation  (3.2)  from the $I_c(E_{11})$ variation of the gravity-dual gravity relations, see equation (2.16),  thus verifying the result of integrating up. 
\par
We observe that equationÊ (3.4) is gauge invariant under the expected gauge transformation for the field $A_{a_1\ldots a_9 , b_1b_2b_3}$, namelyÊ[16]
$$
\delta A_{a_1\ldots a_9 , b_1b_2b_3} = 9\, \partial_{[a_1}\Lambda^{(1)}{}_{a_2\ldots a_9] , b_1b_2b_3}Ê
+ 3\, (\partial_{[b_1 |} \Lambda^{(2)}{}_{a_1\ldots a_9 , | b_2b_3]}Ê
+ {9 \over 7} \,\partial_{[a_1}\Lambda^{(2)}{}_{a_2\ldots a_9][b_1 ,b_2b_3]})
\eqno(3.10)$$
It is important to note that this gauge invariance is not a requirement but instead one finds that the  equation (3.4) which  emerges fromÊ the $E_{11}\otimes_s l_1$ non-linear realisation possess this symmetry. We also observe that  the transformation that occurs in  equation (3.9) can be interpreted as a gauge transformation. 
\par
The general procedure for finding the modulo transformations that occur in the equivalence relations by integrating up expect to be  as follows. For a field that has $r$ blocks of indices we  can find  an equation with $r$ derivatives that varies under $I_c(E_{11})$ transformation into the equations of motion we have already found at lower levels plus higher level terms. Given this equation, which holds exactly, one can integrate it up to find an equation with a lower number of derivatives and in doing so one finds some arbitrary functions which define the equivalence relation. 
The level four field $A_{a_1\ldots a_9, b_1b_2b_3}$ has two blocks of indices and   equation (3.4) which is the equation of motion  has two derivatives and holds exactly. We can also apply this strategy to the  Einstein equation $R_a{}^b=0$ to find the gravity-dual gravity relation given above. It is inevitable that the equations derived in this way  will be an equivalence relations. It would be instructive to carry out the calculations when all the local symmetries of the non-linear realisation are fixed and in particular when the local Lorentz symmetry is used to make the graviton symmetric. We note however that the gravity-dual gravity duality equation would still be an equivalence relation with a transformation in which the parameters are the diffeomorphism of the graviton and the gauge transformation of the dual graviton field;  see equation (5.14) of reference [7]. Indeed the effect of taking this modulo transformation is the same in terms of determining the higher derivative equations  and it might be best to regard this modulo transformation as being the fundamental one. 
\par
We now illustrate  the  second method of ascertaining what is the precise form of the equivalence relations  by carrying out the $I_c(E_{11})$ variations in the context of the $A_{a_1\ldots a_{10} , b_1b_2}$ field which occurs in  equations (3.3), (3.6) and (3.7).Ê From the variation of the gravity-dual gravity relation of equation (2.16) we find equation (3.3) for the field $A_{a_1\ldots a_{10} , b_1b_2}$ and has only one derivative. Varying this latter equation under $I_c(E_{11})$  we find that 
$$
\delta {\cal E}_{c_1...c_{11},\,a_1a_2} ^{(1)}= -\,{60\over 11\cdot 11!}\,\varepsilon_{c_1...c_{11}}\,E^{(1)}_{(a_1|,\,d_1d_2}\,\Lambda_{|a_2)}{}^{d_1d_2}Ê - \varepsilon_{c_1...c_{11}}\,\partial_{(a_1}\,\tilde{\Lambda}_{a_2)},
\eqno(3.11)$$
where 
$$
\partial_{(a_1}\tilde{\Lambda}_{a_2)} = -\,{60\over 11\cdot 11!}\,\left(\partial_{(a_1} h_{\,|d_1d_2|}\,\Lambda_{a_2)}{}^{d_1d_2} + {1\over 20}\,\varepsilon^{d_1...d_{11}}\,\partial_{(a_1} h_{\,|d_1...d_8|,\,a_2)}\,\Lambda_{d_9d_{10}d_{11}}\right).
\eqno(3.12)$$
While from the variation of the dual graviton equation of motion (2.21) we find equation (3.6), which has two derivatives, and its $I_c(E_{11})$ variation is given by 
$$
\delta {\cal E}^{(2)}_{c_1...c_{11},\,a,\,b_1b_2} = {60\over 11\cdot 11!}\,\varepsilon_{c_1...c_{11}}\,\partial_{[b_1|}\,\left(E^{(1)}_{|b_2],\,d_1d_2}\,\Lambda_a{}^{d_1d_2} + E^{(1)}_{a,\,d_1d_2}\,\Lambda_{|b_2]}{}^{d_1d_2}\right)
$$
$$
+ \varepsilon_{c_1...c_{11}}\,\partial_a\,\partial_{[b_1}\,\tilde{\Lambda}_{b_2]},
\eqno(3.13)$$
where 
$\partial_{[b_1}\tilde{\Lambda}_{b_2]} $ is given in equation (3.12), but with antisymmetrisation instead of symmetrisation. 
\par
We observe that these equation do not vary into the field equations that we  already have and we can interpret the additional terms as those required in the equivalence relation. The result is that we we define the equivalence relations 
$$
{\cal E}_{c_1...c_{11},\,a_1a_2} ^{(1)}\dot =0 , \quad {\rm meaning }\quadÊ
{\cal E}_{c_1...c_{11},\,a_1a_2} ^{(1)}- \varepsilon_{c_1...c_{11}}\,\partial_{(a_1}\,\hat {\Lambda}_{a_2)}=0
\eqno(3.14)$$
and 
$$
{\cal E}^{(2)}_{c_1...c_{11},\,a,\,b_1b_2}\dot =0 ,\quad {\rm meaning }\quad {\cal E}^{(2)}_{c_1...c_{11},\,a,\,b_1b_2}+ \varepsilon_{c_1...c_{11}}\,\partial_a\,\partial_{[b_1}\,\hat{\Lambda}_{b_2]}=0
\eqno(3.15)$$
\par
We therefore search for an equation of motion that is independent of  $\hat{\Lambda}_{b}$ by taking  one more derivative.   The result is equation 
(3.7) whose $I_c(E_{11})$ variation is given by 
$$
\delta {\cal E}^{(3)}_{c_1...c_{11},\,a_1a_2,\,b_1b_2} = -\,{60\over 11\cdot 11!}\,\varepsilon_{c_1...c_{11}}\,\partial_{[a_1|}\,\partial_{[b_1|}\,\left( E^{(1)}_{|a_2],\,d_1d_2}\,\Lambda_{|b_2]}{}^{d_1d_2} + E^{(1)}_{|b_2],\,d_1d_2}\,\Lambda_{|a_2]}{}^{d_1d_2}\right),
\eqno(3.16)$$
As it varies into our previous equations of motion we conclude that this equation holds exactly and is a traditional equation rather than an equivalence relation. In carrying out the steps above we have added terms which have higher level derivatives to the equations of motion, namely  ${\cal E}$, but these we have not been shown but can be found in reference [7]. 
\par
We observe that equation (3.7)  is invariant under the gauge transformation [7]
$$
\delta A_{a_1\ldots a_{10},\,b_1b_2} = \partial_{(b_1}\,\Lambda_{|a_1...a_{10}|,\,b_2)} - {10\over 11}\,\partial_{[a_1}\,\Lambda_{a_2...a_{10}]( b_1,\,b_2 )}
$$
$$
+\,\partial_{[a_1}\,\Lambda_{a_2...a_{10}],\,b_1b_2} 
\eqno(3.17)$$
Gauge transformations for the $E_{11}$ theory were proposed in reference [17] and one can verify that the gauge transformations of equation (3.17) and equation (3.10) are precisely of this form. The modulo transformations of equations (3.9), (3.14) and (3.15) are closely related to these gauge transformations. As noted previously the same would be true for the gravity-dual gravity duality relation if one worked in a formalism which fixed completely the local Lorentz symmetry to have the graviton be a symmetric field. It is tempting to assume that the modulo transformations are just the local  transformations given in reference [17]. One could then check if this was consistent with the $I_c(E_{11})$ transformations rather than take the path so far which has been to find the modulo transformations from $I_c(E_{11})$ variations and then see if they are gauge transformations. 
\par
The above procedures explain how to find the equivalence relations as one calculates the equations of motion of the fields level by  level and it would be good to find a systematic procedure. Clearly the transformations that define the equivalence relations are  very closely related to the gauge transformations proposed in reference [17] and it would seem likely that the modulo transformations are indeed just these gauge transformations. As suggested just above assuming this could lead to a systematic method for determining the hierarchy of equations of motion discussed in this paper. We note, however, as the modulo transformations  occur in the $E_{11}$ variations they are gauge transformations  of a  rather specific kind. 
\par
To summarise, the $E_{11}\otimes_s l_1$ non-linear realisation leads to equations of motion for the level four fields that are given above. The $A_{a_1\ldots a_9, b_1b_2b_3}$ field obeys a duality equation with the three form that leads to the equation of motion for the former field, equation (3.8), that is the one required to account for the degrees of freedom given by the irreducible representation of the Poincare group, that is a third rank tensor of the little group SO(9).  The field  
$A_{a_1\ldots a_{10}, b_1b_2}$ obeys equations that lead to no degrees of freedom but it is physical in that  it gives the eleven dimensional origin [7,8] of Romans theory [18].


\medskip
{\bf 4. Discussion of hep-th:1703.01305.}
\medskip
The authors of reference [9] have followed the $E_{11}$ programme [1,2] and constructed the dynamics up to level three at the linearised level. While they have  used the symmetries of  the $E_{11}\otimes_s l_1$ non-linear realisation they have also considered  the    gauge symmetries given in reference [17] as a tool to construct the dynamics. They  encountered problems with describing the gravitational degrees of freedom related to the trace of the spin connection. These  difficulties were previously observed  in reference [13],  however, as was explained in reference [10], and implemented in references [5,6,7],  the equations of motion that follow from the non-linear realisation form a hierarchy with an  increasing number of derivatives  up to and including  equations that have a number of derivatives that is  equal to  the number of blocks of indices on the fields concerned. The equations with less than this number are equivalence relations, that is, they only hold modulo certain transformations while the equations that have the same number of derivatives hold  in a traditional sense and are the equations of motion that we are familiar with. Acting on the equivalence relations  with  derivatives  in a precise way one finds the equations with higher number of derivatives and  the entire system of equations is invariant under the $E_{11}$ symmetries. 
 In this paper we have reviewed the derivation [5,6,7] of the equations of motion  up to and including level four fields. The equations of motion do indeed correctly describe the degrees of freedom of eleven dimensional supergravity including those of the gravity. 
\par
The authors of reference [9] have  added fields to the $E_{11}\otimes_s l_1$ non-linear realisation  in an $E_{11}$ covariant manner using a tensor hierarchy algebra [21] to account for the fields that they thought were missing since they had not realised that the equations with low numbers of derivatives were equivalence relations. In effect  they wish to convert the equivalence relations to be of the usual type of equations of motion.  For the gravity-dual gravity equations they wish to convert the first order gravity-dual gravity duality relation  into a standard equation by adding a nine form field which is related to the trace of the spin connection. This alternative way of proceeding does not affect the equation of motion for a given field with the most number of derivatives and in particular it  does not affect the  equations of motion  for the three form, six form, graviton or the dual graviton as well as the analogous level four equations which were given  in reference [7]. Indeed it also does not affect the full non-linear equations for the fields,  below level four,  which were derived  in references [5,6] and which were found, when suitably truncated,  to be the equations of motion of eleven dimensional supergravity. The  corresponding non-linear equation for the dual graviton will be given in a future publication [15]. 
\par
While there is no need to add new fields to the $E_{11}\otimes_s l_1$ non-linear realisation it could have been advantageous to have a  more familiar system of equations and this may help to find a general method. On the other hand one is, by construction,  adding field which do not appear in the final dynamical equations and so are unphysical. It will be interesting to see how easy is it to add the new fields of reference [9] in comparison with just working with equivalence relations and in particular to see the extension of the results of reference [9] to find  the equations of motion at level four and at the non-linear level. 
\par
An alternative somewhat artificial procedure would be to add fields $ \chi^A$ in the $l_1$ representation and then consider the fields $\hat A_{\underline \alpha}= A_{\underline \alpha}+ (D_{\underline \alpha})_A{}^B\partial_B \chi^A$ where $A_{\underline \alpha}$ are the $E_{11}$ fields. The new fields are obviously invariant under the gauge transformations 
$\delta A_{\underline \alpha}= (D_{\underline \alpha})_A{}^B\partial_B \Lambda^A$ [17] provided one also takes $\delta \chi^A= -\Lambda^A$. Since the modulo transformations of the equivalence relations are closely related to gauge transformations one then has to ensure that the fields $ \chi ^A$ drop out of the equations of motion with the highest number of derivatives. As a result although one now has standard equations the situation is very little different from using the equivalence relations and then removing the modulo transformations by taking derivatives. 
\par
The authors of reference [9] have concentrated on finding  the first order duality equations rather than the hierarchy of equations considered in  references [5,6,7]. In particular, as they did not take account of the equivalence nature of the first order equations,  they had more freedom to derive second order equations and as a result concluded that the $A_{a_1\ldots a_9, b_1b_2b_3}$ field obeys a single trace equation rather than  the correct equation (3.8). Their  approach did not include    deriving  equations of motion which are the second order in derivatives and higher, where required, using $I_c(E_{11})$ variations as was done in reference [7] and in reference [5,6] where  the full non-linear results for the three form and graviton were given. 
\par
The authors of reference [9] noted that the three form equation of motion ${\cal E}_{a_1 , a_2a_3}= 0$ transforms as 
$$
\delta{\cal E}_{a_1  a_2a_3 }\propto \partial^b\partial^{d_1d_2}\partial_{d_1}  \Lambda_{d_2 b a_1 a_2 a_3}
 \eqno(4.1)$$
under the linearised gauge transformations of reference [17]. As a result the authors of reference [9] proposed that the theory should satisfy a section condition. A section condition appeared in Siegel theory [19], more recently called doubled field theory. The one proposed for the $E_{11}$ theory in reference [9] was also discussed  as a possible section condition in reference [20] using BPS arguments. 
\par
Gauge symmetries are usually essential for constructing invariant action using field strengths. For example Maxwell's theory is unique at low energy if one demands that it is invariant under special relativity but  also gauge invariant. However, the situation with $E_{11}$ is rather different;  the equations of motion,  at least at low levels, are determined essentially uniquely by the symmetries of the nonlinear realisation; indeed the result essentially follows from the $E_{11}$ Dynkin diagram. One finds that the equations of motion are automatically invariant under the local  transformations needed for the physical consistency of the theory,  such as diffeomorphisms and  gauge symmetries whose parameters  depend on the usual spacetime As such in the $E_{11}$  theory one does not apparently need to require any gauge transformations to find the equations of motion as they are already determined and have the required  gauge transformations that depend on the usual spacetime.  As such it  would not seem to be  required to impose the additional gauge transformation of reference [17] whose gauge parameters  depend on the higher level coordinates. While the higher level coordinates are crucial for the invariance using the symmetries of the non-linear realisation the physical reason for their presence has yet to be clarified. The strategy of the $E_{11}$ programme to date has been to proceed in a conservative manner by only requiring that the equations of motion be invariant under the $E_{11}$ symmetries.  We note that the modulo transformations used in the equivalence relations that occur in the $E_{11}$ variations are very closely related to gauge transformations but they are very specific field dependent gauge transformations that could well obey special conditions. We note that having to impose a section condition on the gauge transformations is not the same as imposing a section condition on the fields. The authors of reference [9] have calculated to a higher level in the derivatives than previous. It would be interesting to see if  the section condition arise when carrying out the $I_c(E_{11})$ variations to find the equations of motion. 
\par
The non-invariance of equation (4.1) has been know to the author of this  paper for many years. Generalised field strengths in certain lower dimensions were constructed in reference [22] using the gauge transformations of reference [17]. One found that the analogous calculation to that of equation (4.1) did work, that is, the field strengths were invariant  at the level considered. 
This raises the propect that there may be an alternative way of resolving the dilema raised by equation (4.1). 

\medskip
{\bf {Acknowledgements}}
\medskip
We wish to thank Nikolay GromovÊ and Nicolas Boulanger for discussions and Alexander Tumanov for help in preparing this paper. We also wish to thank the STFC for support from Consolidated grant number ST/J002798/1.  
\medskip
{\bf {References}}
\medskip
\item{[1]} P. West, {\it $E_{11}$ and M Theory}, Class. Quant. Ê
Grav.Ê {\bf 18}, (2001) 4443, hep-th/ 0104081.Ê
\item{[2]} P. West, {\it $E_{11}$, SL(32) and Central Charges},
Phys. Lett. {\bf B 575} (2003) 333-342,Ê hep-th/0307098.Ê
\item{[3]} P. West, {\it A brief review of E theory}, Proceedings of Abdus Salam's 90thÊ Birthday meeting, 25-28 January 2016, NTU, Singapore, Editors L. Brink, M. Duff and K. Phua, World Scientific Publishing and IJMPA, {\bf Vol 31}, No 26 (2016) 1630043,Ê arXiv:1609.06863.Ê
\item{[4]} P. West, {\it Very Extended $E_8$ and $A_8$ at low
levels, Gravity and Supergravity}, Class.Quant.Grav. {\bf 20} (2003)
2393, hep-th/0212291.
\item{[5]} A. Tumanov and P. West, {\it E11 must be a symmetry of strings and branes },Ê Phys. Lett. {\bfÊ B759 }Ê(2016),Ê 663, arXiv:1512.01644.Ê
\item{[6]} A. Tumanov and P. West, {\it E11 in 11D}, Phys.Lett. B758 (2016) 278, arXiv:1601.03974.Ê
\item{[7]} A. Tumanov and P. West, $E_{11}$, {\it Romans theory and higher level duality relations}, IJMPA, {\bf Vol 32}, No 26 (2017) 1750023,  arXiv:1611.03369. 
\item{[8]}  A.~Kleinschmidt, I.~Schnakenburg and P.~West, {\it Very-extended Kac-Moody algebras and their interpretation at lowÊ levels}, Class.\  Quant.\  Grav.\ Ê {\bf 21} (2004) 2493 [arXiv:hep-th/0309198].; P.~West, {\it E(11), ten forms and supergravity},Ê JHEP {\bf 0603} (2006) 072,Ê [arXiv:hep-th/0511153].Ê
\item{[9]} G.  Bossard, A.  Kleinschmidt, J.  Palmkvist, C. Pope and  E.  Sezgin, {\it Beyond E11 },  arXiv:1703.01305. 
\item{[10]} P. West, {\it Dual gravity and E11},  arXiv:1411.0920.
\item{[11]} T. Curtright, {\it Generalised Gauge fields}, Phys. Lett. {\bf 165B} (1985) 304. 
\item{[12]} C. Hull, {\it   Strongly Coupled Gravity and Duality}, Nucl.Phys. {\bf B583} (2000) 237, hep-th/0004195. 
\item {[13]} P. West, {\it Generalised Geometry, eleven dimensions
and $E_{11}$}, JHEP 1202 (2012) 018, arXiv:1111.1642. Ê
\item{[14]} X. Bekaert, N. Boulanger and M. Henneaux, {\it Consistent deformations of dual formulations of linearized gravity: A no-go result } 
Class.Quant.Grav. 20 (2003) S417,  arXiv:hep-th/0210278. 
X.  Bekaert, N.  Boulanger and  S.  Cnockaert, {\it No Self-Interaction for Two-Column Massless Fields}, J.Math.Phys. 46 (2005) 012303, arXiv:hep-th/0407102. 
\item{[15]} A. Tumanov and P. West, work in progress 
\item{[16]}Ê N. Boulanger,Ê P, Sundell and P. West, {\it Gauge fields and infinite chains of dualities},Ê , JHEP 1509 (2015) 192,Ê arXiv:1502.07909.Ê
\item{[17]} P. West, {\it Generalised Space-time and Gauge Transformations},
JHEP 1408 (2014) 050,Ê arXiv:1403.6395. Ê
\item{[18]} L. J. Romans, {\sl Massive $N=2A$ supergravity in ten
Ê dimensions}, Phys. Lett. {\bf B 169} (1986) 374.
\item{[19]} W. Siegel, {\it Two vielbein formalism for string inspired axionic gravity},   Phys.Rev. D47 (1993) 5453,  hep-th/9302036; {\it Superspace duality in low-energy superstrings}, Phys.Rev. D48 (1993) 2826-2837, hep-th/9305073; 
{\it Manifest duality in low-energy superstrings},  In *Berkeley 1993, Proceedings, Strings '93* 353,  hep-th/9308133. 
\item{[20]} P. West, {\it Generalised BPS conditions}, Mod.Phys.Lett. A27 (2012) 1250202, arXiv:1208.3397.  
\item{[21]} J. Palmkvist, {\it The tensor hierachy algebra} , J. Math. Phys. 55 (2014) 011701, arXiv:1305.0018. 
\item{[22]} A. Tumanov and P. West, {\it E11 and exceptional field theory },  Int.J.Mod.Phys.A31, (2016) no. 12, 1650066, arXiv1507.08912. 
\item{[22]} N.  Boulanger, S. Cnockaert and  M.  Henneaux, {\it A note on spin-s duality}, JHEP 0306 (2003) 060, hep-th/0306023.


\end